\documentclass[a4paper,fleqn,usenatbib]{mnras}

\newcommand{\msun}{$M_\odot$}
\newcommand{\rsun}{$R_\odot$}
\newcommand{\lsun}{$L_\odot$}

\def\kms{km~s$^{-1}$}

\usepackage[T1]{fontenc}
\usepackage{ae,aecompl}

\usepackage{graphicx}	% Including figure files
\usepackage{amsmath}	% Advanced maths commands
\usepackage{amssymb}	% Extra maths symbols

\title[RS Oph: optical flickering]{The recurrent nova RS Oph  -- simultaneous B and V band observations of the flickering variability}

\author[Zamanov et al.]{
R. K. Zamanov,$^{1}$\thanks{E-mail: rkz@astro.bas.bg (RKZ), glatev@astro.bas.bg (GYL)}
S. Boeva,$^{1}$
G. Y. Latev,$^{1}$
J. Mart\'i,$^{2}$
D. Boneva,$^{3}$ 
B. Spassov,$^{1}$
\newauthor
Y. Nikolov,$^{1}$
M. F. Bode$^{4}$,
S. V. Tsvetkova$^{1}$,
K. A. Stoyanov$^{1}$
\\
\\
$^{1}$ Institute of Astronomy and National Astronomical Observatory, 
       Bulgarian Academy of Sciences,  72 Tsarigradsko Shose, \\
       1784 Sofia, Bulgaria \\
$^{2}$ Departamento de F\'isica (EPSJ), Universidad de Ja\'en, Campus Las Lagunillas, A3-420, 23071, Ja\'en, Spain   \\
$^{3}$ Space Research and Technology Institute,  Bulgarian Academy of Sciences, ul. Akad. G. Bonchev  blok 1, 1113 Sofia, Bulgaria \\
$^{4}$ Office of the Vice Chancellor, Botswana International University of Science and Technology, 
       Private Bag 16, Palapye, Botswana \\
}
\date{Accepted  2018 June 30. Received 2018 June 30 ; in original form 2018 March 7}
\pubyear{2018}
\begin{document}
\label{firstpage}
\pagerange{\pageref{firstpage}--\pageref{lastpage}}
\maketitle

% Abstract of the paper
\begin{abstract} 
We performed 48.6 hours (in 28 nights)  of simultaneous B and V band observations
of the flickering variability of the  recurrent nova RS Oph in quiescence. 
During the time of our observations the brightness of the system varied  between 
$13.2  > B > 11.1 $ and the colour  in the range  $0.86 < B-V < 1.33$.  
We find that  RS~Oph becomes more blue, as it becomes brighter, 
however the hot component becomes more red as it becomes brighter (assuming that the red giant is non-variable). 
During all the runs RS Oph exhibits flickering with amplitude 0.16 - 0.59 mag in B band. 
For the flickering source we find that it has   
colour $-0.14 < B-V < 0.40$,    
temperature in the range  $7200 < T_{fl} < 18900$, 
and average radius  $1.1 < R_{fl}  < 6.7$~\rsun. 
We do not find a correlation between the temperature of the 
flickering and the brightness.  
However, we do find  a strong correlation 
(correlation coefficient 0.81, significance $1.1 \times 10^{-7}$) between B band magnitude
and the average radius of the flickering source --
as the brightness of the system increases  the size of the flickering source also increases. 
The estimated temperature is similar to that of the bright spot of cataclysmic variables.
The persistent presence of flickering indicates
that the white dwarf is actively accreting material for the next outburst.
\end{abstract}

% Select between one and six entries from the list of approved keywords.
% Don't make up new ones.
\begin{keywords}
accretion, accretion discs -- stars: binaries: symbiotic -- novae, cataclysmic variables -- stars: individual: RS Oph

\end{keywords}

%%%%%%%%%%%%%%%%%%%%%%%%%%%%%%%%%%%%%%%%%%%%%%%%%%

%%%%%%%%%%%%%%%%% BODY OF PAPER %%%%%%%%%%%%%%%%%%

\section{Introduction}
RS~Oph (HD 162214) is a symbiotic recurrent nova which exhibits recurrent nova outbursts approximately every 15-20 years.
During the outbursts the brightness rises from an initial value of $V \sim 11$  to $V \approx 6$ magnitude.
Most likely, the recurrent nova outbursts are the result of a thermonuclear  runaway  on the surface 
of the white dwarf (Starrfield 2008), however some authors suggest dwarf nova-like accretion disc instability (King \& Pringle 2009;
Alexander et al. 2011).

RS~Oph contains an M giant mass donor (Dobrzycka \& Kenyon  1994; Anupama \& Miko{\l}ajewska 1999) 
and a massive (1.2-1.4 \msun) carbon-oxygen white dwarf (Miko{\l}ajewska \& Shara 2017). 
The mass transfer mechanism could be via Roche lobe overflow or stellar wind capture, but it is still the subject 
of debates (Somero, Hakala \& Wynn 2017).
Booth, Mohamed \& Podsiadlowski (2016) find that the quiescent mass transfer produces a dense outflow, 
concentrated towards the binary orbital plane, and an accretion disc forms around the white dwarf. 
A photoionization model of the quiescent spectrum indicates the presence of a low-luminosity accretion disc
(Mondal et al. 2018). 

%  GAIA dr2   paralax=0.442  \pm 0.053 --->  d=1/P= 1/0.442   d=2262 pc (2020pc - 2571 pc)

Brandi et al. (2009)  found that the orbital period of the system is 453.6~d and give a mass ratio $q=M_{RG}/M_{WD}=0.59\pm0.05$,
which corresponds to mass of the red giant $M_{RG}$=0.68--0.80~\msun\ and an inclination of the orbit $i = 49^0 - 52^0$. 
The orbit of the system is consistent with it being circular with  period 455.7~d (Fekel et al., 2000). 

The most recent nova outburst of RS~Oph occurred in 2006 February 12 (Narumi et al. 2006; Evans et al. 2008) 
and the multiwavelength observations widen our knowledge on the mechanisms and physics of nova
explosions and formation of planetary nebulae. The optical and radio observations after the outburst confirmed an asymmetric double-ring structure of the nova remnant 
(O'Brien et al. 2006; Bode et al. 2007; Rupen, Mioduszewski \& Sokoloski 2008). 
The X-ray observations show that the nova interacts with the dense circumstellar medium leading to deceleration of 
the material (Sokoloski et al. 2006; Bode et al. 2006). 
The ejecta have low mass -- M$_{ej}$ $\sim$ 10$^{-7}$ -- 10$^{-6}$~\msun, 
and high velocity -- $\upsilon_{ej}$ $\gtrsim$ 4000~\kms\ (Bode et al. 2006; Sokoloski et al. 2006; Vaytet et al. 2011).
%RS~Oph exhibits recurrent nova outbursts approximately every 20 years. So far, outbursts have been recorded in 1898, 1933, 1958, 1967,
%1985 and most recently in 2006 (Evans et al., 2008; Narumi et al. 2006). According to Schaefer (2010), two outbursts in 1907 and 1945 were missed
%when RS~Oph  was  aligned  with  the  Sun. Most likely, the recurrent nova outbursts are the result of a thermonuclear  runaway  on the surface 
%of the white dwarf (Starrfield 2008), however some authors suggest dwarf nova-like accretion disc instability (King \& Pringle 2009;
%Alexander et al. 2011).

The flickering is seemingly irregular with light variations on timescales of
a few minutes with amplitude of a few$\times0.1$ magnitudes. 
The flickering of RS~Oph has been observed by Walker (1977), Bruch (1980), Gromadzki et al. (2006) and others.
The study of flickering in any interacting binary system is important because it allows us to probe 
what is going on in the inner accretion disc regions. The release of luminosity here follows variations 
at larger accretion disc radii on time-scales of the order, or larger than, the local viscous time-scale.
The variations are expected to be of flicker-noise type according to $\alpha$-based accretion disc 
models pioneered by Lyubarskii (1997). 
Improving our knowledge of flickering in RS Oph enables us to widen the 
exploration of the parameter space of accretion discs around mildly compact objects
such as its white dwarf companion. In this context, our RS~Oph observations contribute to a complementary view of
flickering sources beyond previous studies, which have been more focused on
black holes in X-ray binary and Active Galactic Nuclei systems.
With its unusual flickering properties among symbiotic
stars, RS Oph provides an excellent alternative workbench
for theorists to put to the test their accretion disc models, and
help to understand phenomena such as the widely observed
rms-flux relation (see Scaringi et al. 2012). 

Here, we report quasi-simultaneous observations of the flickering variability of 
the recurrent nova RS Oph in two optical bands -- B and V and
analyze the colour changes, temperature and radius of the flickering source
and their response to the brightness variations. 

% Our aim is to check what happens in the flickering source - 
% does it change temperature, radius luminosity when the brightness of the system
% increases / decreases ?  (???)

%-----------------------------------------------------------------------------   
 \begin{figure*}    
   \vspace{9.2cm}     
   \includegraphics{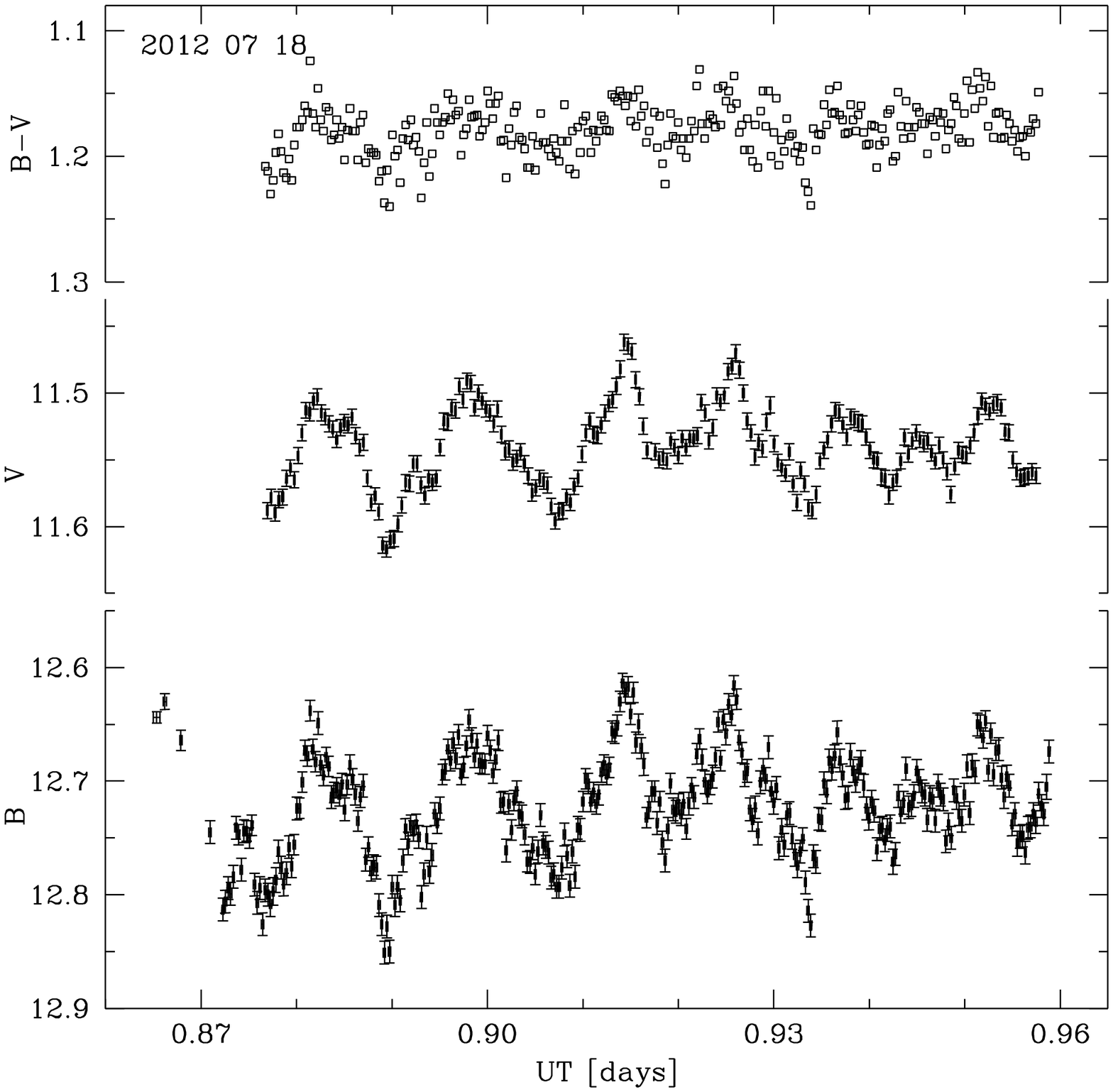}      
   \includegraphics{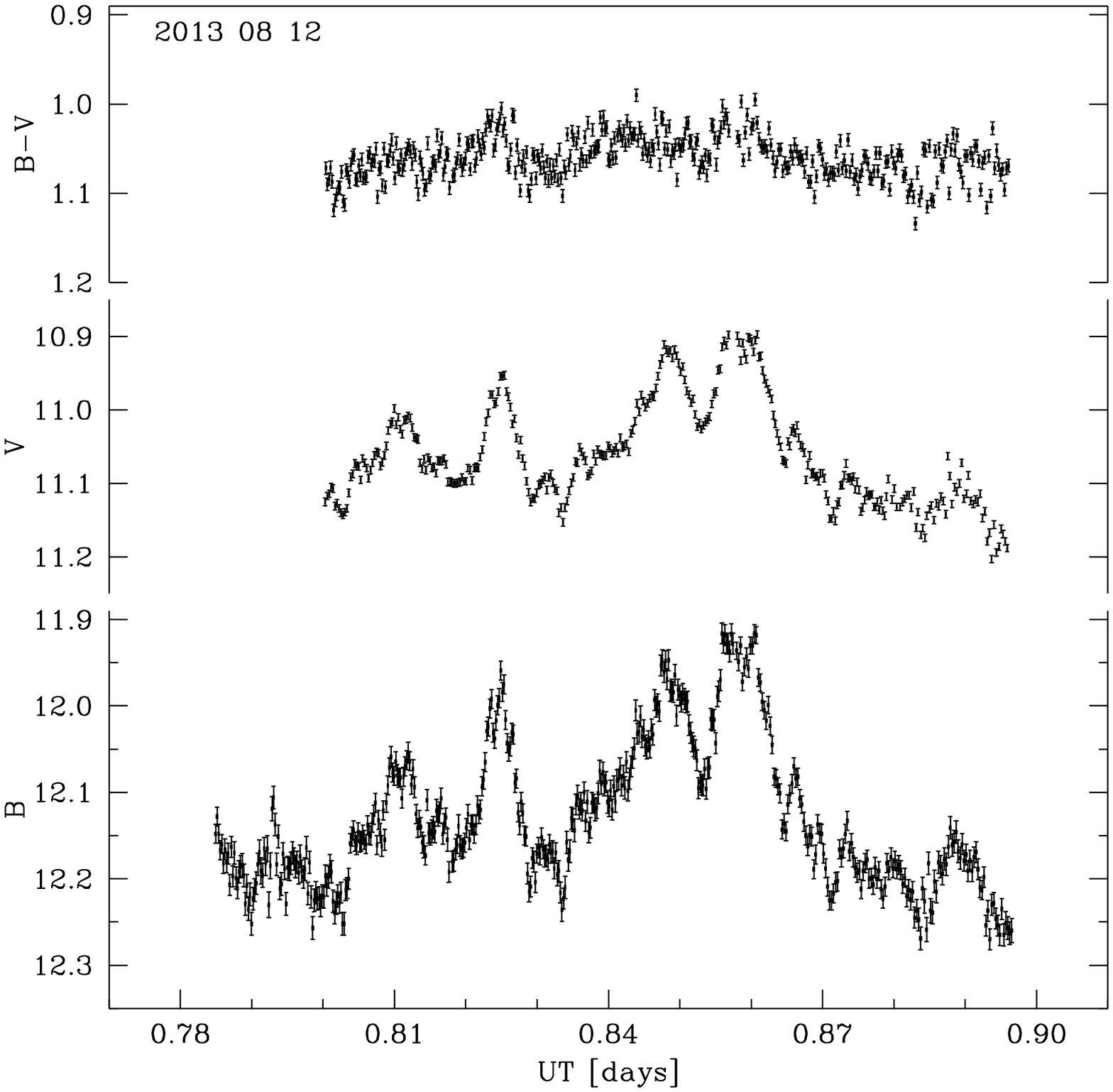}      
   \caption[]{Example of our observations - 
   light curves of RS Oph in B and V bands together with the calculated B-V colour.   }
   \label{fig.ex}      
 \end{figure*}	     
%------------------------------------------------------------------------------ 

\section{Observations}

The observations were performed with five telescopes equipped with CCD cameras: 
\begin{itemize}
  \item the 2.0 m telescope of the National Astronomical Observatory (NAO) Rozhen, Bulgaria  
  (Bonev \& Dimitrov 2010)  
  \item the 50/70 cm Schmidt telescope of NAO Rozhen    
  \item the 60 cm telescope of NAO Rozhen  
  \item the 60 cm telescope of the Belogradchik Observatory, Bulgaria (Strigachev \& Bachev 2011)  
  \item the 41~cm telescope of the University of Ja\'en, Spain 
(Mart{\'{\i}}, Luque-Escamilla, \&  Garc\'{\i}a-Hern\'andez 2017).  
% , 0.42 arcsec px$^{-1}$). 
\end{itemize}
The data reduction was done with IRAF (Tody 1993) following standard recipes for processing of CCD images and aperture photometry. 
A few comparison stars from the list of Henden \& Munari (2006) have been used. 
On Fig.~\ref{fig.ex} two examples of our data are given -- the observed variability in B and V bands 
is plotted together with the calculated B-V colour.
The typical photometric errors are 0.006~mag in B band and 0.005~mag in V band.

We have 28 nights with simultaneous observations in B and V bands 
during the period July 2008  - September 2017.  The $B-V$ colour is calculated for 2749 points in total.
During our observations the brightness of RS~Oph was: \\ 
$11.121  \le  B    \le  13.208 $, \\
$10.093  \le  V    \le  11.908 $,  \\
$ 0.864  \le  B-V  \le   1.333 $,\\
with  mean B = 12.204,  mean V = 11.106,  mean B-V = 1.098. 
The peak-to-peak amplitude of the flickering in B band is in the range  0.16 - 0.59 mag.
% The highest amplitude --- ... the lowest amplitude - duration of the run ....

The journal of observations is given in Table~\ref{tab1}. 
In the table are given as follows:
the date (in format YYYYMMDD), band, the telescope, number of data points over which B-V colour is calculated, 
average, minimum and  maximum magnitude in the corresponding band, 
dereddened colour of the flickering source $(B-V)_{01}$ and  $(B-V)_{02}$,  
T$_1$  and  T$_2$ - temperature of the flickering source, 
R$_{fl}$  and  R$_2$ - radius the flickering source.
Details of the calculation are given in Sect.~\ref{flickering} below.
%-----------------------------------------------------------------------------   
 \begin{figure}    
   \vspace{6.0cm}     
   \includegraphics{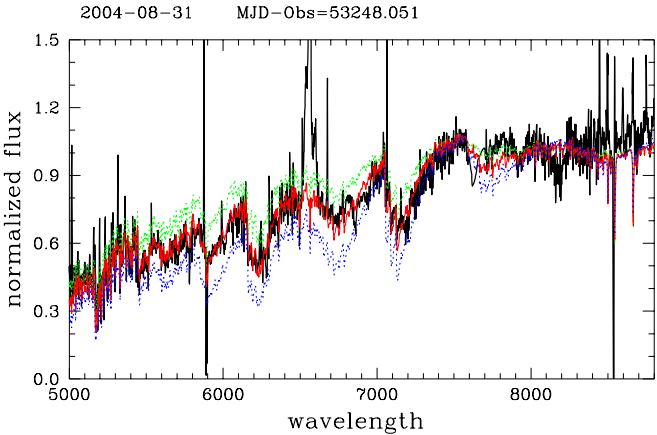}      
   \caption[]{The spectrum of RS Oph is  denoted by the black line. The spectrum is corrected for interstellar extinction 
   using E(B-V)=0.69   and the contribution of the hot continuum is subtracted. 
   The green, red and blue lines are template spectra 
   of  M1III, M2III and M3III giants, respectively,  from  Zhong et al. (2015). }
   \label{fig.rg}      
 \end{figure}	     
%------------------------------------------------------------------------------ 

\section{Interstellar extinction and red giant contribution}

\subsection{Interstellar reddening}

We have nine high resolution spectra obtained with the FEROS spectrograph attached to the 2.2m telescope 
ESO La Silla and the echelle spectrograph ESPERO at the 2.0m telescope of NAO Rozhen. 
We compared these spectra with spectra of a few red giants 
with similar spectral type and measured the equivalent widths of four interstellar features
clearly  visible in the high resolution spectra. 
Using the equivalent width of  KI$\lambda 7699$ and 
the calibration by  Munari \& Zwitter (1997), we find $0.60<E(B-V)<0.83$.
Using the diffuse interstellar bands (DIBs) and  results of  Puspitarini,  Lallement \& Chen (2013),
we derive  for DIB $\lambda 6613$  $0.59<E(B-V)<0.80$,
for DIB 5780  $0.57<E(B-V)<0.69$,
for DIB	5797  $0.65<E(B-V)<0.78$.
Taking into account the individual errors,
we estimate $E(B-V)=0.69 \pm 0.07$. This is in agreement with  
(1) the value $E(B-V)=0.73 \pm 0.10$  given in  Snijders (1987) on the basis of UV spectra 
and to (2) the Galactic dust reddening for a line of sight to RS Oph (NASA/IPAC 
IRSA: Galactic Reddening and Extinction Calculator) $0.59 \le E(B-V) \le 0.69$, 
which should be considered as an upper limit for the interstellar reddening.
Hereafter we  use  $E(B-V)=0.69 \pm 0.07$. 

\subsection{Red giant}
\label{subsect.redgiant}

On basis of  the light curves of RS~Oph 
after the 2006 outburst (see Sect.~3 in Zamanov et al. 2015)
we estimated that the brightness of the red giant is $m_V \approx 12.26$. 
To check this value we assume that the ratio between the variable and non-variable part of the 
hot component contribution 
to the brightness is equal in U and V bands. 
The flickering of RS Oph is clearly visible in all UBVRI bands with amplitude increasing to the blue. 
Using data from Zamanov et al. (2010, 2015), 
we estimate the magnitude of the red giant in V band to be 
$m_V = 12.19 \pm 0.10$. This value (although based on an assumption) 
is in agreement with $m_V \approx 12.26$ above. 

Shenavrin, Taranova \& Nadzhip (2011) found M2pe for the spectral type of the red giant from IR photometry.
Mondal et al. (2018) from the absorption band indices derived
the spectral type to be M2-M3. 
 Skopal (2015a) found that the radiation from the giant corresponds to its 
effective temperature $T_{eff} = 3800 - 4000$ K, and the radius $R_g = (61 - 55) (d/1.6 \; {\rm kpc})$~\rsun .
Comparing our spectra with  spectra of red giants
(Houdashelt et al. 2000;  Bagnulo et al. 2003; Zhong et al. 2015), 
we estimate a similar spectral type M2~III.
In Fig.\ref{fig.rg} we plot one of our spectra of RS Oph. 
The spectrum is 
corrected for the interstellar extinction using E(B-V)=0.69, 
the contribution of the hot continuum is subtracted,
and the spectrum is normalized at 8350 \AA. The emission lines are not removed. 
We compared the spectrum with templates of M giants by Zhong et al. (2015).
The red line is the template  spectrum of an  M2III giant  from  Zhong et al. (2015), for which we achieved 
the best agreement. 

An M2III star is expected to have an effective temperature 3695  --  3750~K,
and radius  57.8 --  60.5~\rsun\ (van Belle et al. 1999). 
 For such a giant and solar metallicity models  Houdashelt et al. (2000) give 
$T_{eff} =  3740$~K,  $\log g = 0.81$,  $B-V= 1.629$, $V-I = 1.927$.
We adopt for the brightness of the red giant in the optical bands 
$m_B=14.660$  and $m_V=12.261$. 
The typical $1 \sigma$ error of these magnitudes is about $\pm 0.05$ mag.

\section{Variability in B and V bands}

In Fig.\ref{fig.mm} we plot B versus V band magnitude.
In the left  panel (Fig.~\ref{fig.mm}a) are the observed magnitudes of RS Oph. 
In the right panel (Fig.~\ref{fig.mm}b) are the dereddened magnitudes of
the hot component (e.g. the red giant contribution is subtracted using the magnitudes given
in Sect. \ref{subsect.redgiant}).  
% The magnitudes of the hot component are calculated from the observed,
% subtracting the red giant contribution (see Sect. \ref{subsect.redgiant}). 

\subsection{B-V colour}

\begin{figure*}    
   \vspace{7.0cm}     
   \includegraphics{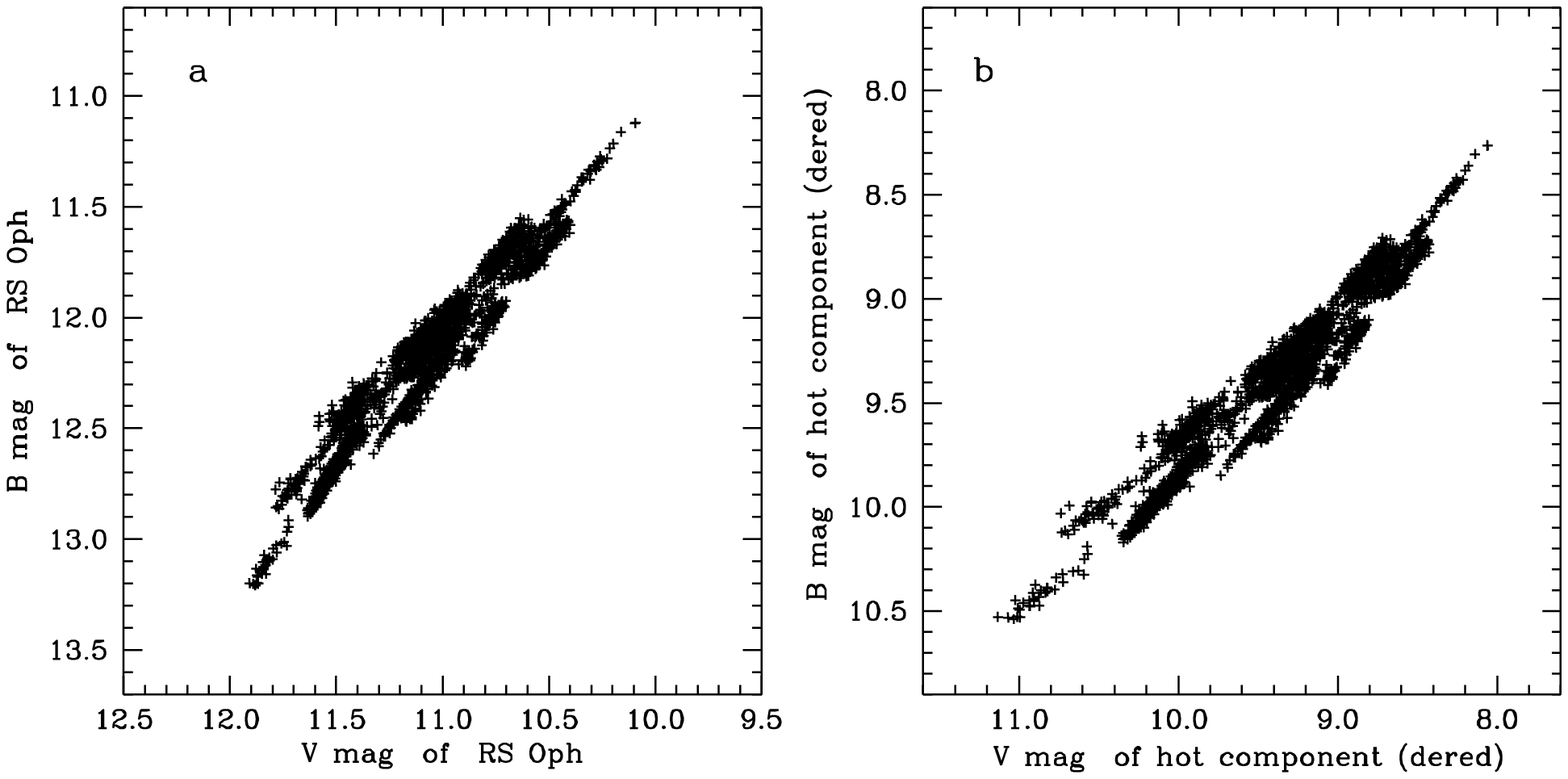} 
   \caption[]{B versus V band magnitude:   a) observed,  b) calculated for the hot component. } 
   \label{fig.mm} 
   \vspace{7.0cm}      
   \includegraphics{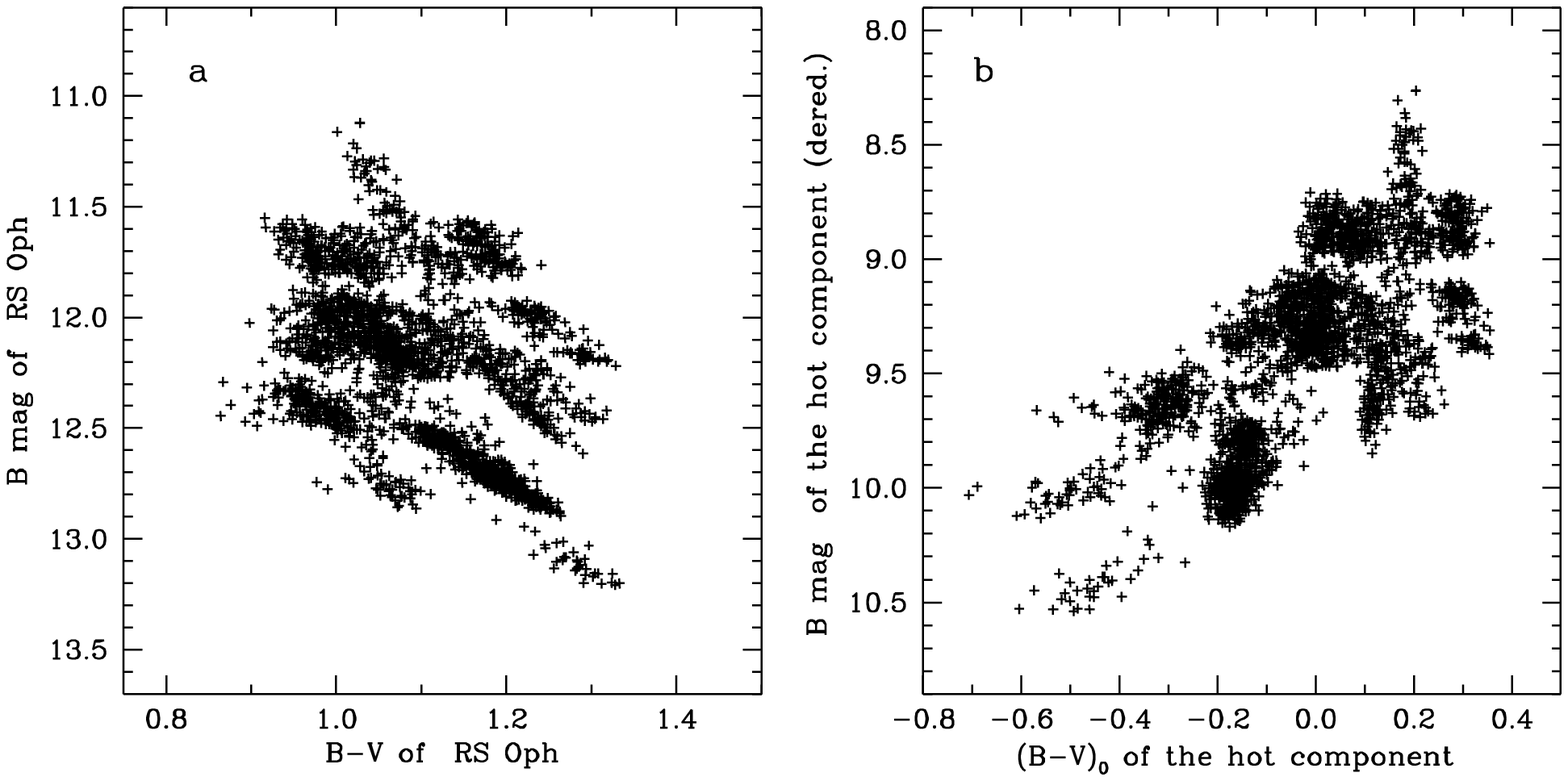}  
   \caption[]{Colour magnitude diagram:  a) observed,  b) calculated for the hot component.  } 
   \vspace{7.0cm} 
   \label{fig.cm2}        
   \includegraphics{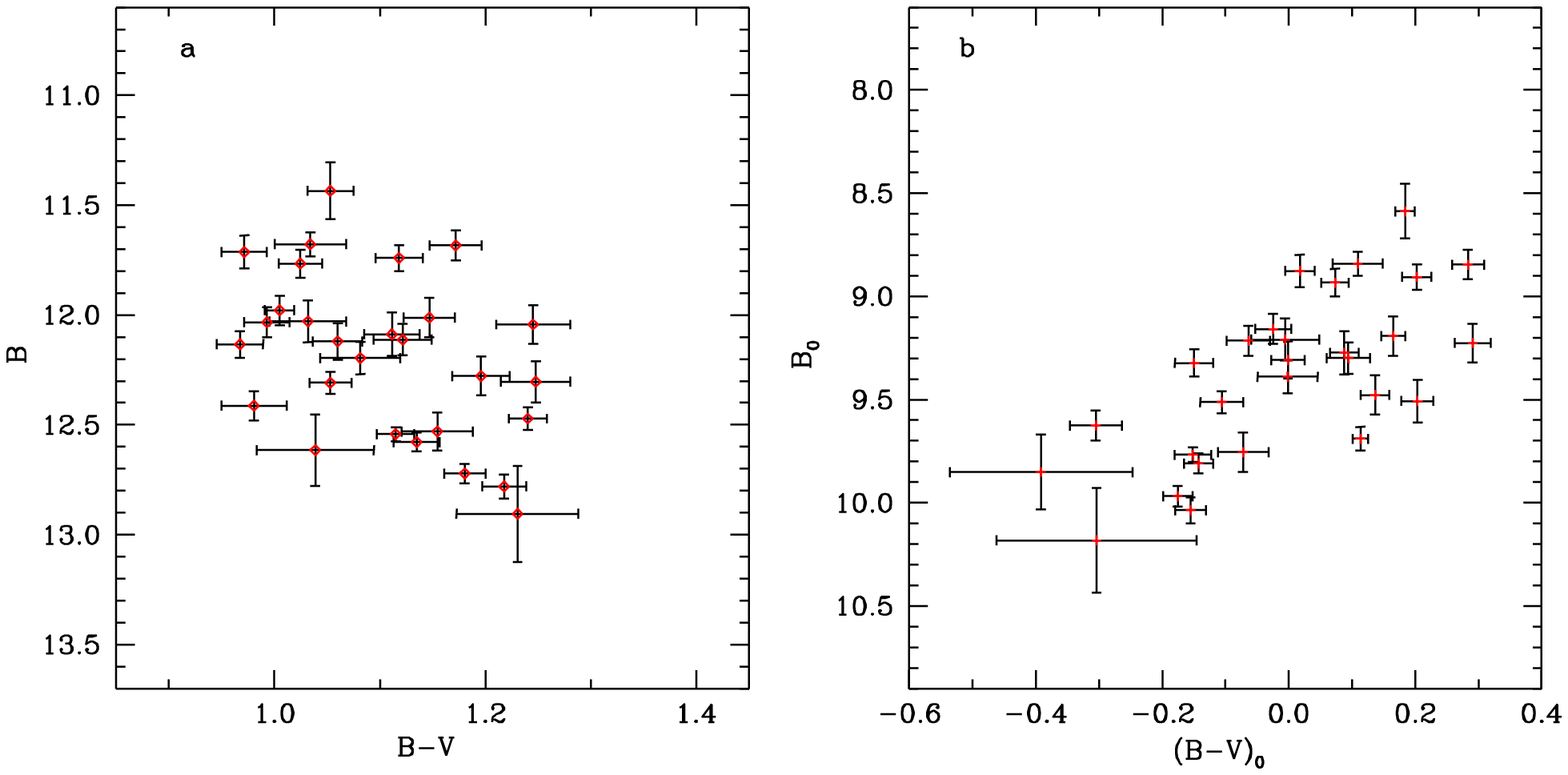}      
   \caption[]{Mean colour magnitude diagram, each point represent one run: 
     a) observed,  b) calculated for the hot component. }
   \label{fig.cm3}      
 \end{figure*}

In Fig.\ref{fig.cm2}, we plot colour magnitude diagrams. 
The colour-magnitude diagram is quite different from that of the cataclysmic variable AE Aqr (Zamanov et al. 2017). 
In the case of AE Aqr all the data occupy a well defined strip, 
here such a strip is not visible, although the observations from each night are placed on a strip on the diagram.
This indicates  that the flickering behaviour in the case RS Oph is more complicated and/or with a different mechanism. 

In Fig.\ref{fig.cm3}, we plot the calculated mean values for each night (one night -- one point).
The error bars correspond to the standard deviation of the run. 
% The left panel is the  observed, right panel is the hot component (red giant contribution subtracted).
It is seen  that the system becomes bluer as it becomes brighter, however  
the hot component becomes redder as it gets brighter.
There is a  correlation between the mean colour and magnitude of the hot component.
The Pearson correlation coefficient is 0.70, Spearman's (rho) rank correlation -- 0.69,
the statistical significance  $p\mbox{--}value = 5.5 \times 10^{-5}$. This indicates that the hot component 
becomes redder as it gets brighter. 

%-----------------------------------------------------------------------------   
%------------------------------------------------------------------------------ 

% \subsection{B-V colour of the hot source and flickering}
% Our results point to that the flickering itself does not change the B-V colour of the hot source ???

% 0.8 to 1.0 very strong relationship
% 0.6 to 0.8  strong relationship
% 0.4 to 0.6  moderate relationship
% 0.2 to 0.4  weak relationship
% 0 to 0.2    weak or no relationship

% 0 - 0.2 - very poor or very weak
% 0.2 - 0.4  - poor or weak
% 0.4 - 0.65 - fair or moderate
% 0.65 - 0.85 - strong or high
% 0.85 - 1.0  - very strong / high 
% 1.0  - perfect 

\section{Flickering light source}
\label{flickering}
In his remarkable paper, Bruch (1992) proposed that the light curve of
a white dwarf with flickering  can be separated into two parts -- constant light
and variable (flickering) source. 
% We assume that all the variability 
% in each night is due to flickering. In these suppositions the flickering 
% light source is considered 100\% modulated.  
Following his assumptions,
we calculate the flux of the flickering light source 
as $F_{\rm fl1}=F_{\rm av}-F_{\rm min}$, where $F_{\rm av}$ is the average flux 
during the run and $F_{\rm min}$ is the minimum flux during the run
(corrected for the typical error of the observations).

An extension of the method is proposed by Nelson et al. (2011), 
who suggests to use the $F_{\rm fl2}=F_{\rm max}-F_{\rm min}$, where $F_{\rm max}$ 
is the maximum flux during the run. 

Practically, the method of Bruch (1992) refers to the average luminosity of the flickering source, 
while that of Nelson et al. (2011) -- to its maximal luminosity. 
$F_{\rm fl1}$ and $F_{\rm fl2}$  have been calculated for each band, using the values 
given in Table~1 and 
the calibration for a zero magnitude star $ F_0 (B) =6.293 \times 10^{-9}$  erg cm$^{-2}$ s$^{-1}$ \AA$^{-1}$,    $\lambda_{eff}(B)=4378.12$~\AA, 
$F_0 (V) = 3.575 \times 10^{-9}$  erg cm$^{-2}$ s$^{-1}$ \AA$^{-1}$ and  $\lambda_{eff}(V)=5466.11$~\AA\ as given in Spanish virtual observatory 
Filter Profile Service  (Rodrigo et al. 2018, see also Bessel 1979).  

It is worth noting that while the calculated colours of the hot component depend on
the assumed red giant brightness, the parameters of the flickering source are independent on the 
the red giant parameters. 

Using method of Bruch (1992), 
we find that in B band the flickering light source  contributes about 
13\%  of the average flux of the system,  with    $ 0.06  \le  F_{\rm fl1} / F_{\rm av} \le 0.24$.
In V band its average contribution is 11\%,  with    $ 0.05  \le  F_{\rm fl1} / F_{\rm av} \le 0.19$.

Using method of Nelson et al (2011), 
we find that in B band the flickering light source  contributes about 
25\%  of the maximal flux of the system,  with   $0.13   \le  F_{\rm fl2} / F_{\rm max} \le 0.42 $.
In V band its is about 21\%,  with  $  0.10  \le  F_{\rm fl2} / F_{\rm max} \le 0.36$.

From the amplitude - flux relation (rms-flux relation), we expect that
the luminosity of the flickering source will increase as the brightness increases. 
However, it is not a priori clear which parameter --  temperature or radius (or both)  increases.

\subsection{B-V colour and temperature of the flickering source}

The calculated de-reddened  colours of the flickering light source are given in Table~\ref{tab1},  
where $(B-V)_{01}$ is calculated using $F_{av}$ and $F_{min}$, while  $(B-V)_{02}$ is calculated using $F_{max}$
and  $F_{min}$. Typical error is $\pm 0.05$ mag. 

% They are calculated using the calibration $ F_0 (B) =6.293e-9$  erg cm$^{-2}$ s$^{-1}$ \AA$^{-1}$    $\lambda_{eff}=4378.12$~\AA, 
% $F_0 (V) = 3.575e-9   $  erg cm$^{-2}$ s$^{-1}$ \AA$^{-1}$    $\lambda_{eff}=5466.11$~\AA as given in Spanish virtual observatory 
% The SVO Filter Profile Service  (Rodrigo et al.2018). 

\begin{figure}    
   \vspace{7.8cm}     
   \includegraphics{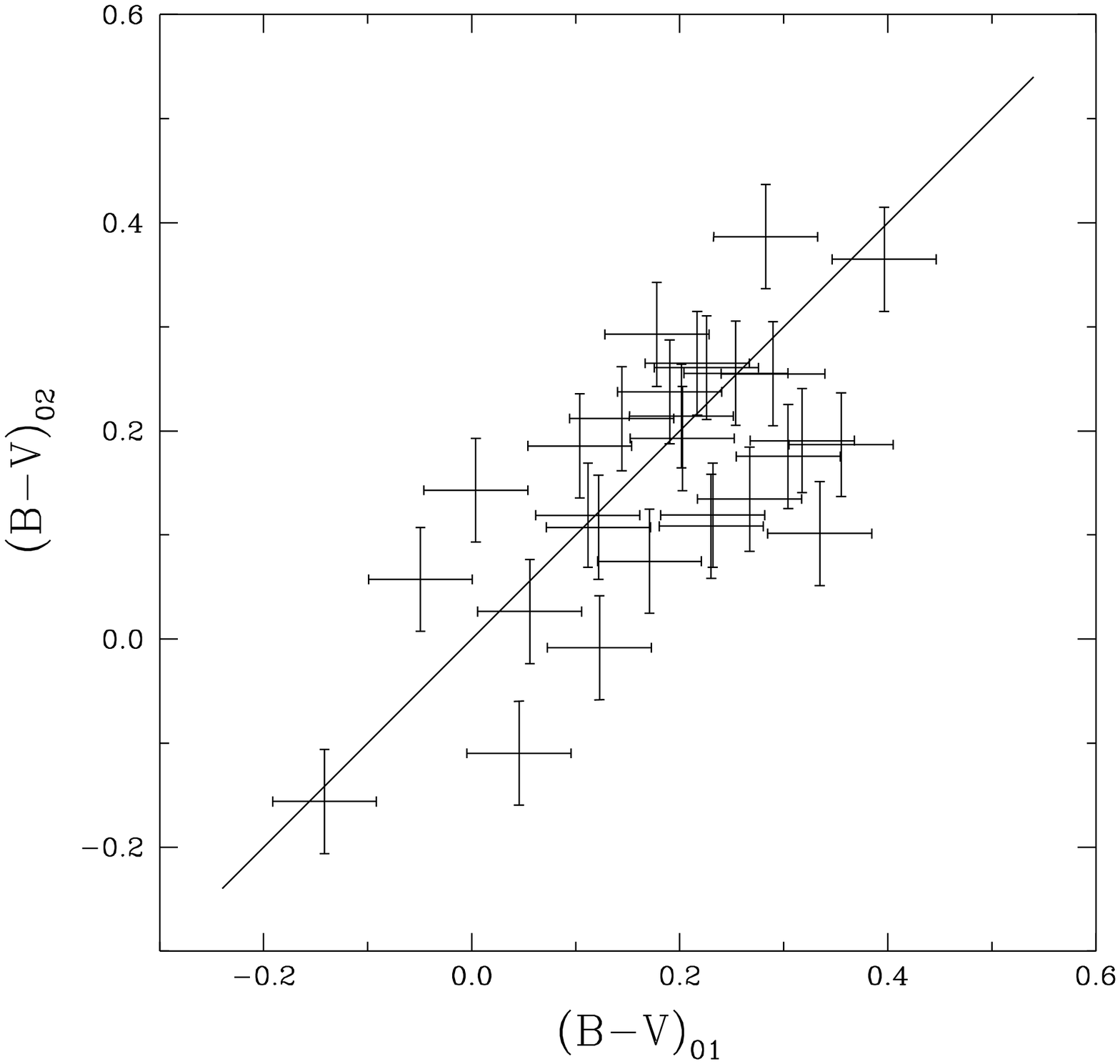}      
   \caption[]{$(B-V)_{02}$ versus $(B-V)_{01}$ --
    there is no systematic shift between the two methods. See text for details.}
 %   $(B-V)_0$ colour of the flickering source versus the average B magnitude. 
 %   The squares represent  $(B-V)_{01}$ calculated by method of 
 %   Bruch (1992), the plus signs -- $(B-V)_{02}$  derived following Nelson et al. (2010).  
 %   There is no tendency 
 %   for the flickering source to become redder or bluer, when the brightness changes. }
   \label{fig.41}      
   \vspace{5.4cm}     
   \includegraphics{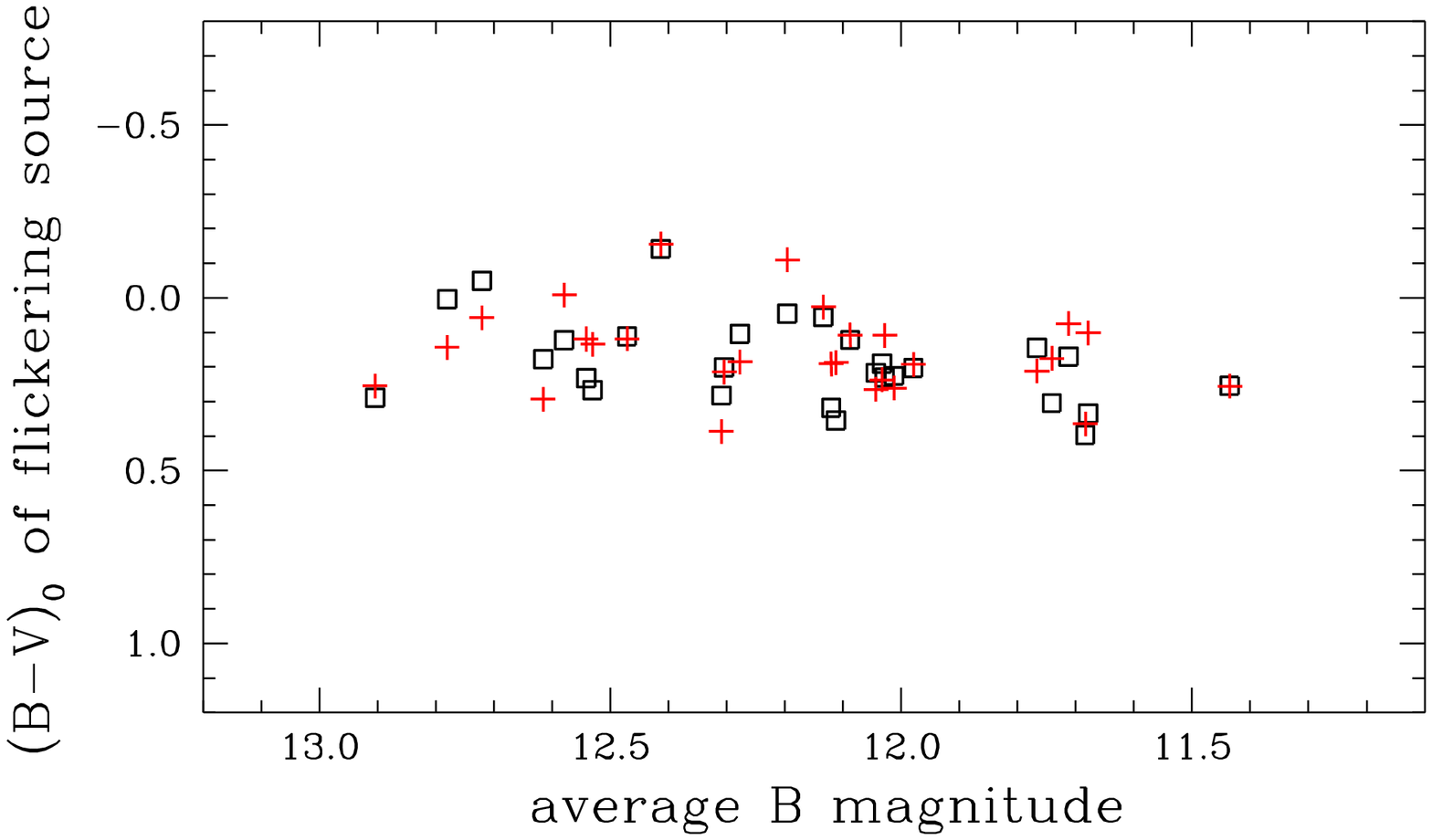}      
   \caption[]{ $(B-V)_0$ colour of the flickering source versus the average B magnitude. 
    The squares represent  $(B-V)_{01}$ calculated by method of 
    Bruch (1992), the plus signs -- $(B-V)_{02}$  derived following Nelson et al. (2011).  
    There is no tendency 
    for the flickering source to become redder or bluer when the brightness changes. }
    \label{fig.42}      
 \end{figure}

In Fig.\ref{fig.41}  we plot $(B-V)_{02}$  versus $(B-V)_{01}$.  
The solid line represents $(B-V)_{02} = (B-V)_{01}$. 
% It is visible that both methods give similar results and there is not a systematic shift.
To  check for a systematic shift between the two methods we performed linear least-squares 
approximation in one-dimension $(y = a + b x)$, when both x and y data have errors.
We obtain $a= 0.027 \pm 0.017$ and $ b = 1.007 \pm 0.093$. 
A Kolmogorov-Smirnov test gives Kolmogorov-Smirnov statistic 0.18 and  significance level 0.72.
It means  that both methods give similar results and there is not a systematic shift.

The average difference between them is $\approx  0.08$ mag,
which is comparable with the accuracy of our estimations. 
In Fig.\ref{fig.42}, we plot $(B-V)_0$ versus the average B band magnitude. We do not detect a correlation
between the colour of the flickering source and brightness. 

\subsection{Temperature of the flickering source}
\label{temp.fl}

We calculate the temperature of the flickering source using its dereddened colours 
and the colours of the black body (Strai\v zys, Sud\v zius  \& Kuriliene 1976).
$T_1$ is calculated using $(B-V)_{01}$, and $T_2$ is calculated using  $(B-V)_{02}$. 
The values are given in Table~\ref{tab1}. 
The two methods give similar results for
the  temperature of the flickering source as well as for $(B-V)_{0}$. 
The average  values are $T_1 = 9835 \pm 2400$ and  $T_2 = 10306 \pm 2693$. The slightly larger scatter 
of the values calculated following Nelson et al. (2011) is due to the fact that 
$F_{\rm av}$  is  calculated more precisely than $F_{\rm max}$.

%-----------------------------------------------------------------------------   
%------------------------------------------------------------------------------ 

%-----------------------------------------------------------------------------   
 \begin{figure}    
   \vspace{11.5cm}     
   \includegraphics{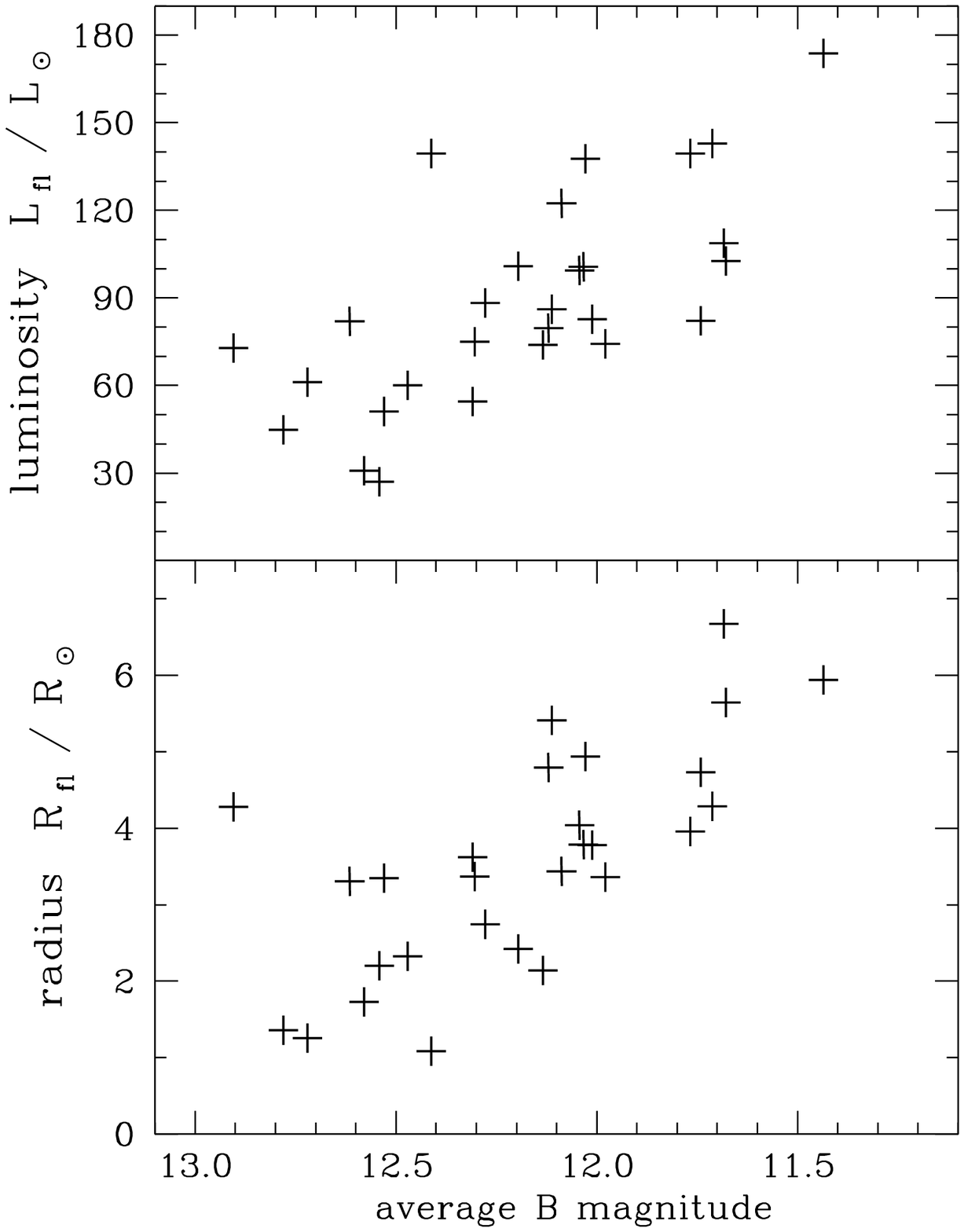}      
   \caption[]{Luminosity and radius of the flickering source versus the average B band magnitude. 
    There is a well defined tendency 
    for the flickering source to become brighter and bigger when the brightness of the system increases.
    %(It is well visible that the flickering source become larger when the brightness increases).
    }
   \label{fig.Rfl}      
 \end{figure}	     
%------------------------------------------------------------------------------ 

% \subsection{B-V versus B-V of the hot component Radius of the flickering source}
% Please plot   $(B-V)_{01}$  versus $(B-V)_0$ of the hot component Radius of the flickering source

\subsection{Radius of the flickering source}
\label{R.fl.1}

The radius of the flickering source  $R_{fl}$ is calculated 
using the derived temperature (Sect.\ref{temp.fl}), B band magnitude and assuming  that it is spherically symmetric. 
We obtain  $0.92  <  R_{fl} < 5.6$~\rsun.

In Fig.\ref{fig.Rfl} we plot $R_{fl}$ versus the average B band magnitude. It is seen that 
$R_{fl}$ increases  when the brightness of the system increases. 
When we use all 28 points we  find moderate to strong correlation 
with  Pearson correlation coefficient   0.70, Spearman's (rho) rank correlation   0.73,  
and statistical  significance  $p\mbox{--}value \approx 9 \times 10^{-6}$. 

The most deviant point is 
%  point that deviates significantly from the correlation.
the run 2012 August 15,  the same run that also deviates  from the 
rms-flux relation  (see Sect. 6.1 in Zamanov et al. 2015).
When we remove this deviating point  
(due to the  low brightness and high amplitude variability, i.e. $N_{pts} =27$), 
we obtained strong and highly 
significant correlation between the radius of the flickering source and average B band magnitude 
of the hot component with Pearson correlation coefficient 0.81,  
Spearman's (rho) rank correlation  0.83, statistical significance of the correlation 
$p\mbox{--}value \approx 1.1 \times 10^{-7}$. 

We do not detect significant correlation between the brightness and 
the radius of the flickering source calculated with $F_{max}$. 

\subsection{Luminosity  of the flickering source}
\label{L.fl.1}

The luminosity of the flickering source  $L_{fl}$ is calculated 
using the derived temperature and radius given in Table~\ref{tab1}  
and assuming  that it is spherically symmetric:  $L_{fl} = 4 \pi R_{fl}^2 \sigma T_1^4$, 
where  $\sigma$ is the Stefan-Boltzmann constant. 
We obtain  27~\lsun ~ $<  L_{fl} < 173$~\lsun.

\section{Discussion}

%RS Oph has 8 recorded outbursts between 1898 and 2016, with time interval between 
%two consecutive outbursts between 8.6 and 26.6 years (Schaefer 2010; Adamakis et al. 2011). 
% RS~Oph is one of only 11 symbiotic stars with detected flickering. 
% The flickering detection has been reported by Walker (1977) and later 
% it was studied in details (Bruch 1980, 1992; 
% Dobrzycka, Kenyon \& Milone 1996; Sokoloski, Bildsten \& Ho 2001; Simon, Hudec \& Hroch 2004; Gromadzki et al., 2006; Zamanov et al., 2010, 2015).
% RS~Oph is one of only 11 among more than 200 symbiotic stars with detected flickering. 
%These outbursts are a result of thermonuclear runaway on the surface of the WD. 

The flickering variability is typical  for the accreting white dwarfs in cataclysmic variables and is 
considered an observational proof of accretion onto a white dwarf (Sokoloski \& Bildsten 2010).
It is a relatively rare phenomenon for symbiotic stars. 
Among more than 200 symbiotic stars known in our Galaxy, flickering is observed only in 11 objects. 
The last two detected in recent years V648~Car (Angeloni et al. 2013) and EF~Aql (Zamanov et al. 2017).
The flickering of RS Oph disappeared after the 2006 outburst (Zamanov et al. 2006), 
reappeared by day 241 of the outburst (Worters et al. 2007),  
and is visible in all our observations obtained since then to now. 
This indicates that the white dwarf is accumulating material for the next outburst.

Outbursts of RS~Oph have been recorded in 1898, 1933, 1958, 1967,
1985 and most recently in 2006 (Evans et al., 2008; Narumi et al. 2006). 
According to Schaefer (2010), two outbursts in 1907 and 1945 were missed
when RS~Oph  was  aligned  with  the  Sun. 
The time interval between two consecutive outbursts is between 8.6 and 26.6 years 
(Schaefer 2010; Adamakis et al. 2011).
The last one took place in 2006, which means that 
the next one can be expected between the time of writing  and 2032. 
We are now well into the observed historic inter-outburst period and 
we should be alert for the next eruption. 

Changes in the emission from H$\alpha$ line are detected on time-scales as short as 2 min
(Worters \& Rushton  2014). This time scale is similar to that of the flickering variability
and indicates that the flickering is partly re-processed in the nebula. 
The variability observed in  H$\alpha$  could be due to changes in the
photoionization of the nebula  linked to the flickering activity. 
Given the discussion in Sect.~6.2 of Sokoloski, Bildsten \& Ho (2001) about the difficulty of
producing rapid variability from the nebular emission, we 
expect  the rapidly variable component of RS Oph 
to reflect the physical origin of the variations, 
for example in the accretion disc (or boundary layer, or hot spot), 
and not be dominated by nebular features.

In the following paragraphs we discuss the implication of our results for 
three possible sites of origin of the flickering - the hot spot (e.g. Smak 1971), 
structures in the accretion disc (e.g.  Baptista,  Borges  \& Oliveira 2016) and the boundary layer (e.g. Bruch \& Duschl 1993).

\subsection{Typical time of the flickering}
\label{RKepl}

The wavelet analysis of the flickering of RS~Oph performed 
by Kundra, Hric \& G{\'a}lis (2010) and Kundra \& Hric (2014) found two different sources of
flickering, the first one with amplitude of 0.1 magnitude and frequency 60-100 cycles
per day and the second one with amplitude of 0.6 magnitude and frequency less than
60 cycles per day (with both parameters
varying from night to night). Similar frequencies (17 - 144 cycles/day) are also visible in our data (Georgiev et al. 2018). 
Simon, Hudec \& Hroch (2004) using observations with the Optical Monitoring Camera onboard of INTEGRAL satellite,  
found that the typical frequency was 30-50 cycles/day (i.e. period of 48-29 min) during 
the observations in 2003. Semiregular oscillations with period $82 \pm 2$~min on one night (about 17 cycles per day)
have been also detected by  Dobrzycka, Kenyon \& Milone (1996).

Let us suppose that these quasi-periods correspond to
the Keplerian period in the accretion disc around the WD:
\begin{equation}
   R_{Kepl} = \left( \frac{ G  M_{WD} }{4 \pi ^2  \nu^2} \right)   ^{1/3},  
 \label{eq.R.kepl}
\end{equation}
where $G$ is the gravitational constant, $R_{Kepl}$ is the distance from the WD, and $\nu$  is the frequency of the variability. 
For a $1.4$~\msun\ WD the frequencies correspond to  $R_{Kepl} \approx 0.2 - 0.5$~\rsun.
These values are 5-10 times smaller than the calculated (in Sect.~\ref{R.fl.1}) 
average radius of the flickering source. 
However they are similar to the radii estimated in Sect.~\ref{Tdisc}.

%where $G$ is the gravitational constant,  $R_{Kepl}$ is the distance from WD, $\nu$  is the frequency of the variability. 
%For an $1.4$~\msun\  WD the frequencies correspond to $R_{Kepl} \approx 0.2 - 0.5$~\rsun. 
%These values are 5-10 times smaller than the calculated (in Sect.~\ref{R.fl.1}) 
%average radius of the flickering source. 
%However they are similar to the radii estimated in Sect.~\ref{Tdisc}. %

% Let we suppose that these quasi-periods somehow correspond to the free-falling time from distance $r$: 
% \begin{equation}
%   t_{ff} =  \frac{r^2}{ \sqrt { 6 G  M_{WD} } }. 
% \label{eq.tff}
% \end{equation}
% For an $1.4$~\msun\  WD they correspond to $r  \approx  $~\rsun. 

% These values are smaller but similar to the average radius of the flickering source. 

% how to calculate free falling time ?????  
% \begin{equation}
%   t_{ff} = \left( \frac{ G  M_{WD}  }{4 \pi ^2  \nu^2 } \right)   ^{1/?}.  ?  
% \label{eq.R.ff}
% \end{equation}

% \subsection{Radius of the flickering source}
% If the flickering is due to vortexes - it means that the average size of the
% vortexes  increases as the mass accretion rate increases. %
% If the flickering is due to boundary layer - it means that the average size of the boundary layer 
% increases as the brightness increases.  
%  If the flickering is due to a bright spot  - it means that the average size of the bright spot 
% increases as the brightness increases.  

\subsection{ Bright spot }

The temperature and the size of the bright spot 
are derived for a few 
cataclysmic variables. 
For OY Car, Wood et al (1989) calculated
temperature in the range 8600 -- 15000~K; 
Zhang \& Robinson (1987) for U Gem - $T = 11600 \pm 500$ K;
Robinson, Nather \& Patterson (1978) give $T = 16000$~K 
for the bright spot in WZ Sge.  
For IP~Peg three estimates exist:   
Marsh (1988) -- $T = 12000 \pm 1000$~K,  Ribeiro et al. (2007) -- 6000-10000 K, 
Copperwheat et al. (2010) -- 7000 - 13000 K.
The temperature of the optical flickering
source of RS Oph is in the range $7000 < T_{fl} < 18000$~K (see Sect.4.1), which is similar to
the temperature of the bright spot for the CVs. 

The bright spot is produced by the impact of the stream on the outer parts of the
accretion disc. In case of Roche-lobe overflow this stream is coming from the inner Lagrangian point $L_1$.
If the red giant in RS~Oph does not fill its Roche lobe, the white dwarf accretes material from its wind. 
In this case accretion cone and accretion wake will be formed [see  Fig.~1 of Jackson (1975) or
Fig.~4 of  Ahmad, Chapman \& Kondo (1983)]. 
The stream formed in the accretion wake should be similar to that formed from Roche-lobe overflow.  

The luminosity of the bright spot is approximately (Shu 1976;  Elsworth \& James 1982): 
\begin{equation}
 L_{bs} \approx  \frac{1}{2}  v_\perp ^2 \dot M_{acc},  
\label{eq.hs}
\end{equation}
where $\dot M_{acc}$ is the mass accretion rate and $v_\perp$ is the inward component 
of the stream's velocity at the impact with  outer disc edge. 
Eq.~\ref{eq.hs}  indicates that when the mass accretion rate
increases, the luminosity of the hot spot also must increase.
In addition, our results (Sect.\ref{R.fl.1}) indicate that when the mass accretion
rate increases the radius of the bright spot (if it is the source
of flickering) also increases, while its temperature remains
almost constant. In this case the quasi-periods (Sect. \ref{RKepl})
are not connected with the Keplerian rotation but most likely with
fragmentation and/or variability in the stream.

% Eq.\ref{eq.hs} indicates that when the mass accretion rate increases the luminosity 
% of the hot spot also must increase. 
% Our results (Sect.\ref{R.fl.1}) indicate that when the mass accretion rate increases 
% the radius of the  bright spot (if it is the source of flickering) also increases, 
% while  its temperature remains almost constant. In this case the quasi-periods (Sect. \ref{RKepl})
% are  not connected with the Keplerian rotation but likely with fragmentation and/or variability in the stream. 

\subsection{Temperature in the accretion disc}
\label{Tdisc}

The timescales of changes of the overall structure of the accretion disc 
(e.g.  the mass transfer rate, angular momentum transport, global spiral structure formation)
are longer compared to the local fluctuating processes in the flow that 
are responsible for the flickering activity. In this way 
the dynamical time scale variability of the flickering light source do not
change the overall structure of the accretion disc.
Considering the entire disc structure, 
the temperature in the disc can be approximated with 
the radial temperature profile of a steady-state accretion disc (e.g.  Frank, King \& Raine 2002):
% Considering the entire disc structure, 
% as an approximation the radial temperature profile
% (e.g.  Frank, King, Raine 2002)  of a steady-state accretion disc
% (e.g.  Frank, King, Raine 2002) is:
% of the stationary (or steady state?) 
% disc and we describe the radial effective temperature, by applying the equations of Frank, King, Raine (2002) : 
% The radial temperature profile of a steady-state accretion disc is:
\begin{equation}
T_{eff}^4=\frac{3 G \dot M_{acc} M_{wd}}{8 \pi \sigma R^3} 
\left[ 1-\left(\frac{R_{wd}}{R}\right)^{1/2} \right] ,
\end{equation}
$R$ is the radial distance from the WD.
We assume   $ M_{wd} = 1.35$~\msun. In the standard model an white dwarf with mass  1.2-1.4 \msun\ is expected to have radius
0.006 - 0.002 \rsun\  (e.g. Magano et al.  2017). 
The average mass accretion rate in quiescence is of about  $2.3 \times  10^{-7}$ ~\msun~yr$^{-1}$ (Skopal 2015b).
Bearing in mind the variability of the hot component in B band (see Fig.~\ref{fig.cm3}b) it implies that
the mass accretion rate in quiescence  varies approximately
from $1 \times 10^{-7}$  to $5.5 \times  10^{-7}$~\msun~yr$^{-1}$. 

Using the parameters for RS~Oph, 
a temperature  $7500 \le T_{fl} \le 14500$~K 
(the temperature of the flickering light source as derived in  
Sect.~\ref{temp.fl}) should be achieved at a distance 
$R\approx 0.5-2.0$ \rsun\ from the WD.
If the vortexes are 1000 K hotter than their
surroundings than this distance is  $R\approx 0.6-2.6$ \rsun\ from the WD.
These values are similar to the Keplerian radii corresponding to the quasi-periods in the flickering light curves (see Sect. \ref{RKepl}).
If the accretion disc itself (vortexes, blobs or other structures in the disc, e.g. Bisikalo et al. 2001) 
is the place for the origin of the flickering of RS~Oph, then it 
comes from $R \sim 1$~\rsun\ from the WD. 

% Using the considerations in the Sect.\ref{R.fl.1}, Sect.\ref{RKepl} and in this section, the flickering source is most probably the bright spot.
%Bearing in mind that this radius is considerably smaller than
%that calculated in Sect.~\ref{R.fl.1}, 
%the flickering source more likely is the bright spot.

\subsection{X-ray emission from the boundary layer} 
RS Oph is an X-ray faint recurrent nova. Practically X-ray observations point to 
a "missing boundary layer" (Mukai 2008).
However the flickering source of RS Oph is considerably brighter in the visible and near infrared (V, R and I bands)
than the flickering source of T CrB, which is a X-ray bright source (Mukai et al. 2013). 
For example in R and I bands RS Oph has flickering with an amplitude
larger than 0.20 mag, while in T CrB the flickering in R and I bands is with amplitude less than 0.03 mag.
In V band  RS Oph has flickering with an amplitude larger than 0.30 mag, while T CrB 
it is   $\le 0.05$ mag (e.g. Zamanov et al. 2010, 2016).

If the boundary layer is completely optically thick, this can explain why RS Oph is X-ray faint. 
In this case the radius of the flickering source measured in the optical bands (Fig.~\ref{fig.Rfl})
could represent the radius up to which the hard X-rays generated from the boundary layer 
are effectively processed by the inner parts of the accretion disc and the accretion disc corona. 

%  (reprocessed by the optically (and X-ray) thick accretion disc). 

\section{Conclusions}
We report our observations of the flickering variability of 
the recurrent nova RS Oph, simultaneously in B and V bands. 
We estimate interstellar reddening towards RS~Oph  $E(B-V)=0.69 \pm 0.07$
and spectral type of the giant M2~III.
Subtracting the red giant contribution, we find that  dereddened colour  $(B-V)_0$ 
of the hot component 
changes from -0.4 to 0.3 and  it becomes redder when it gets brighter. 

For the flickering light source in RS~Oph, we estimate average $(B-V)_0 = 0.18 \pm 0.12$,  
$T_{fl} =  9800 \pm 2400$~K, which is similar to the temperature 
of the bright spot in CVs. Its average radius is $R_{fl} = 3.6 \pm 1.4$~\rsun, 
and average luminosity  $L_{fl} = 89 \pm 35$~L$_\odot$.
% We do not detect relationship between the brightness and temperature of the flickering. 
% However We do find that there is a strong correlation between the brightness and 
% the radius of the flickering source. 
When the brightness of the hot component increases 
the temperature of flickering source  remains approximately constant
and the radius of flickering source increases.   

If the flickering of RS Oph is coming from a bright spot our results indicate that 
when the brightness increases the size of the spot also increases. If the flickering 
comes from the accretion disc, it is probably generated at distance  $\sim 1$~\rsun\ from the white dwarf.

% The possible sites/mechanisms  for the origin of the  flickering are briefly discussed. 
In any case, the richness of the photometric behaviour reported in this manuscript represents a challenging dataset
for further theoretical studies.

%
% The last numbered section should briefly summarise what has been done, and describe 
% the final conclusions which the authors draw from their work.
%

\section*{Acknowledgements}

This work was partly supported by the Program for career development of young scientists, 
Bulgarian Academy of Sciences (DFNP 15-5/24.07.2017),
by grants  DN 08/1~13.12.2016, DN 12/13~12.12.2017 (Bulgarian National Science Fund), 
and  AYA2016-76012-C3-3-P from the Spanish Ministerio de Econom\'ia y Competitividad (MINECO).
We are grateful to an anonymous referee whose comments and suggestions helped to improve the original manuscript.

%%%%%%%%%%%%%%%%%%%%%%%%%%%%%%%%%%%%%%%%%%%%%%%%%%

%%%%%%%%%%%%%%%%%%%% REFERENCES %%%%%%%%%%%%%%%%%%

% The best way to enter references is to use BibTeX:

%\bibliographystyle{mnras}
%\bibliography{example} % if your bibtex file is called example.bib

% Alternatively you could enter them by hand, like this:
% This method is tedious and prone to error if you have lots of references

\appendix

%-------------------------------------------------------------------
\begin{table*} %[!hb]
\small
\caption{Simultaneous CCD observations of RS Oph in B and V bands and the calculated parameters of the flickering source. 
$(B-V)_{01}$, T$_1$, and  R$_{fl}$ - are dereddened colour, temperature and radius of the  flickering source calculated following Bruch (1992), 
$(B-V)_{02}$, T$_2$, and  R$_{2}$ -  following Nelson et al. (2011) -- see Sect.~\ref{flickering}. }
\begin{center}
\begin{tabular}{llc r ccc rrrrrrrrrr}
\hline
DATE     & band &  telescope & $N_{B-V}$ &  average min max     & $(B-V)_{01}$ & T$_1$  & R$_{fl}$ & $(B-V)_{02}$ & T$_2$ & R$_2$ && \\
         &      &            &           & [mag] [mag] [mag]    &  	       &  [K]   & [\rsun] &          & [K]   & [\rsun]&& \\
\hline
20080706 & B  &   70cm Sch  &	 49 &  12.4711 12.380  12.615  & 0.1117  &	 10515   &	 2.322 &       0.1189  &       10384   &       3.096 &&\\	  
         & V  &   2.0m Roz  &	    &  11.2308 11.168  11.325  & 	 &		 &	       &	       &	       &	     &&\\
%20090614 & B  &   60cm Roz  &	 69 &  11.7237 11.605  11.855  & -0.0939 &	 16328   &	 1.713 &       0.2439  &       8701    &       6.238 &&\\
%         & V  &   60cm Bel  &	    &  10.6935 10.557  10.780  & 	 &		 &	       &	       &	       &	     &&\\										
20090615 & B  &   60cm Roz  &	 76 &  11.7666 11.597  11.951  & 0.1442  &	 9948	 &	 3.954 &       0.2118  &       9102    &       6.686 &&\\	  
         & V  &   60cm Bel  &	    &  10.7420 10.581  10.895  & 	 &		 &	       &	       &	       &	     &&\\								  
20090721 & B  &   60cm Roz  &	 57 &  12.0280 11.876  12.277  & 0.2302  &	 8872	 &	 4.937 &       0.1083  &       10576   &       4.773 &&\\	  
         & V  &   60cm Bel  &	    &  10.9963 10.891  11.219  & 	 &		 &	       &	       &	       &	     &&\\					
20090723 & B  &   70cm Sch  &	 38 &  11.9789 11.857  12.099  & 0.2025  &	 9219	 &	 3.360 &       0.1927  &       9341    &       4.793 &&\\	  
         & V  &   2.0m Roz  &	    &  10.9740 10.865  11.082  & 	 &		 &	       &	       &	       &	     &&\\							  
20100430 & B  &   60cm Roz  &	 62 &  11.7409 11.603  11.848  & 0.3043  &	 7964	 &	 4.732 &       0.1754  &       9558    &       5.199 &&\\	  
         & V  &   60cm Roz  &	    &  10.6227 10.521  10.718  & 	 &		 &	       &	       &	       &	     &&\\									  
20100501 & B  &   60cm Roz  &	 66 &  11.4349 11.121  11.612  & 0.2539  &	 8576	 &	 5.936 &       0.2554  &       8558    &     10.7083 &&\\ 
         & V  &   60cm Roz  &	    &  10.3817 10.093  10.541  & 	 &		 &	       &	       &	       &	     &&\\								
20120427 & B  &   60cm Roz  &	 58 &  12.1117 11.934  12.271  & 0.3551  &	 7541	 &	 5.408 &       0.1868  &       9415    &       5.271 &&\\	  
         & V  &   60cm Bel  &	    &  10.9899 10.863  11.138  & 	 &		 &	       &	       &	       &	     &&\\									  
20120613 & B  &   60cm Roz  &	 33 &  12.5301 12.380  12.671  & 0.2671  &	 8411	 &	 3.349 &       0.1345  &       10100   &       3.564 &&\\	  
         & V  &   60cm Roz  &	    &  11.3755 11.274  11.492  & 	 &		 &	       &	       &	       &	     &&\\									  
20120718 & B  &   70cm Sch  &	200 &  12.7207 12.617  12.850  & -0.0493 &	 14379   &	 1.254 &       0.0573  &       11504   &       2.351 &&\\	  
         & V  &   2.0m Roz  &	    &  11.5400 11.462  11.617  & 	 &		 &	       &	       &	       &	     &&\\								 
20120721 & B  &   60cm Bel  &	199 &  12.4126 12.201  12.552  & -0.1415 &	 18972   &	 1.086 &       -0.1561 &       19783   &       1.736 &&\\	  
	 & V  &   60cm Bel  &	    &  11.4321 11.288  11.524  & 	 &		 &	       &	       &	       &	     &&\\							  
20120815 & B  &   60cm Roz  &	 63 &  12.9043 12.618  13.208  & 0.2897  &	 8129	 &	 4.277 &       0.2549  &       8564    &       5.761 &&\\	  
	 & V  &   70cm Sch  &	    &  11.6736 11.452  11.908  & 	 &		 &	       &	       &	       &	     &&\\									  
20120816 & B  &   2.0m Roz  &	185 &  12.7805 12.666  12.897  & 0.0039  &	 12783   &	 1.358 &       0.1430  &       9962    &       2.887 &&\\	  
	 & V  &   70cm Sch  &	    &  11.5624 11.472  11.633  & 	 &		 &	       &	       &	       &	     &&\\								 
20130702 & B  &   60cm Roz  &	 46 &  12.1955 11.982  12.368  & 0.0455  &	 11718   &	 2.422 &       -0.1098 &       17211   &       2.321 &&\\	  
	 & V  &   60cm Roz  &	    &  11.1140 10.988  11.237  & 	 &		 &	       &	       &	       &	     &&\\									  
20130710 & B  &   70cm Sch  &	127 &  12.0329 11.887  12.207  & 0.1903  &	 9371	 &	 3.784 &       0.2375  &       8781    &       5.997 &&\\	  
	 & V  &   70cm Sch  &	    &  11.0399 10.896  11.196  & 	 &		 &	       &	       &	       &	     &&\\									  
20130812 & B  &   70cm Sch  &	335 &  12.1202 11.916  12.270  & 0.3177  &	 7852	 &	 4.796 &       0.1906  &       9368    &       5.476 &&\\	  
	 & V  &   2.0m Roz  &	    &  11.0603 10.897  11.203  & 	 &		 &	       &	       &	       &	     &&\\							 
20130813 & B  & 70Sch+60Roz &	 91 &  12.5788 12.476  12.659  & 0.1228  &	 10313   &	 1.728 &       -0.0084 &       13152   &       1.855 &&\\	  
	 & V  & 2.0 m Roz   &	    &  11.4438 11.381  11.503  & 	 &		 &	       &	       &	       &	     &&\\								    
20130906 & B  &   60cm Bel  &	 39 &  12.0118 11.848  12.152  & 0.2258  &	 8928	 &	 3.779 &       0.2608  &       8490    &       6.376 &&\\	  
	 & V  &   60cm Bel  &	    &  10.8648 10.723  10.977  & 	 &		 &	       &	       &	       &	     &&\\								    
20140621 & B  &   60cm Bel  &	 65 &  12.5414 12.459  12.614  & 0.2318  &	 8852	 &	 2.199 &       0.1190  &       10382   &       2.469 &&\\	  
	 & V  &   60cm Bel  &	    &  11.4264 11.369  11.487  & 	 &		 &	       &	       &	       &	     &&\\							 
20140622 & B  &   60cm Bel  &	104 &  12.6152 12.347  12.864  & 0.1780  &	 9525	 &	 3.305 &       0.2928  &       8090    &       6.958 &&\\	  
	 & V  &   60cm Bel  &	    &  11.5765 11.303  11.786  & 	 &		 &	       &	       &	       &	     &&\\							
20140729 & B  &   70cm Sch  &	 32 &  12.3089 12.228  12.430  & 0.2826  &	 8218	 &	 3.619 &       0.3866  &       7278    &       6.191 &&\\	  
	 & V  &   70cm Sch  &	    &  11.2557 11.163  11.368  & 	 &		 &	       &	       &	       &	     &&\\							
20140831 & B  &   70cm Sch  &	 54 &  12.0877 11.950  12.303  & 0.1218  &	 10331   &	 3.434 &       0.1073  &       10595   &       4.361 &&\\	  
	 & V  &   70cm Sch  &	    &  10.9762 10.873  11.136  & 	 &		 &	       &	       &	       &	     &&\\							
20160726 & B  &   70cm Sch  &	 63 &  12.3041 12.151  12.471  & 0.2016  &	 9230	 &	 3.367 &       0.2142  &       9072    &       4.993 &&\\	  
	 & V  &   2.0m Roz  &	    &  11.0561 10.941  11.174  & 	 &		 &	       &	       &	       &	     &&\\							
20160728 & B  &   70cm Sch  &	 92 &  12.2771 12.084  12.455  & 0.1039  &	 10656   &	 2.742 &       0.1855  &       9431    &       5.107 &&\\	  
	 & V  &   70cm Sch  &	    &  11.0810 10.926  11.201  & 	 &		 &	       &	       &	       &	     &&\\							 
20170329 & B  &   70cm Sch  &	116 &  12.0432 11.925  12.219  & 0.2167  &	 9041	 &	 4.041 &       0.2650  &       8438    &       6.140 &&\\	  
	 & V  &   70cm Sch  &	    &  10.7978 10.701  10.924  & 	 &		 &	       &	       &	       &	     &&\\								
20170528 & B  &   60cm Bel  &	 88 &  12.1335 11.960  12.252  & 0.0558  &	 11531   &	 2.142 &       0.0264  &       12108   &       3.259 &&\\	  
	 & V  &   60cm Bel  &	    &  11.1661 11.028  11.262  & 	 &		 &	       &	       &	       &	     &&\\								
20170626 & B  &   60cm Bel  &	144 &  11.6834 11.557  11.814  & 0.3965  &	 7196	 &	 6.672 &       0.3649  &       7459    &       8.890 &&\\	  
	 & V  &   60cm Bel  &	    &  10.5115 10.400  10.632  & 	 &		 &	       &	       &	       &	     &&\\								 
20170719 & B  &   41cm Ja\'en &	133 &  11.7123 11.551  11.895  & 0.1709  &	 9614	 &	 4.284 &       0.0747  &       11187   &       4.755 &&\\	  
	 & V  &   41cm Ja\'en &	    &  10.7409 10.617  10.905  & 	 &		 &	       &	       &	       &	     &&\\						       
20170904 & B  &   41cm Ja\'en &	 65 &  11.6785 11.557  11.805  & 0.3347  &	 7711	 &	 5.645 &       0.1013  &       10704   &       4.409 &&\\	  
	 & V  &   41cm Ja\'en &	    &  10.6443 10.567  10.770  &         &               &             &	       &	       &	     &&\\
\end{tabular}																      
\end{center}																               
\label{tab1}
\end{table*}
%-------------------------------------------------------------------
	      							                        	        				     
\end{document}